# LOW LEVEL RF SYSTEM OF THE LIGHT PROTON THERAPY LINAC

D. Soriano Guillen†, G. De Michele, S. Benedetti, Y. Ivanisenko, M. Cerv, AVO-ADAM, Meyrin, Switzerland

*Abstract*

The LIGHT (Linac for Image-Guided Hadron Therapy) project was initiated to develop a modular proton accelerator delivering beam with energies up to 230 MeV for cancer therapy. The machine consists of three different kinds of accelerating structures: RFQ (Radio-Frequency Quadrupole), SCDTL (Side Coupled Drift Tube Linac) and CCL (Coupled Cavity Linac). These accelerating structures operate at 750 MHz (RFQ) and 3 GHz (SCDTL, CCL). The accelerator RF signals are generated, distributed, and controlled by a Low-Level RF (LLRF) system. The LIGHT LLRF system is based on a commercially available solution from Instrumentation Technologies with project specific customization. This LLRF system features high amplitude and phase stability, monitoring of the RF signals from the RF network and the accelerating structures at 200 Hz, RF pulse shaping over real-time interface integrated, RF breakdown detection, and thermal resonance frequency correction feedback. The LLRF system control is integrated in a Front-End Controller (FEC) which connects it to the LIGHT control system. In this contribution we present the main features of the AVO LLRF system, its operation and performance.

## INTRODUCTION

AVO-ADAM designed and is currently commissioning the LIGHT (Linac for Image Guided Hadron Therapy) proton cancer therapy LINAC [1], which is a modular normal conducting RF accelerator fed by 4 Inductive Output Tubes (IOTs) and 13 klystrons grouped in 14 power stations. At each power station the RF power can be modulated independently every pulse. The pulse repetition rate is 200 Hz allowing accurate dose delivery within tumour volume and longitudinal layer switching on a pulse-to-pulse basis, given the low emittance of the proton beam. These features of the LIGHT LINAC are key to have image-guided adaptative radiation therapy with protons [1]. The modular structure of the LIGHT system consists of:

- A proton source injecting 40 keV protons with currents up to 300 uA and pulses up to 20 us at 200 Hz repetition rate.
- an RFQ (Radio Frequency Quadrupole) with a resonant frequency of 749.48 MHz accelerating the protons up to 5 MeV. This is the fourth sub-harmonic of the 2997.92 MHz LINAC frequency. The RFQ is fed by an IOT powering system driven by the first LLRF unit. Several signals are monitored in the LLRF from the RFQ system: 4 probe signals from the RFQ cavity and 4 pairs of directional couplers (forward and reflected power) in the RF network.
- Four SCDTL (Side Coupled Drift Tube LINAC) structures powered by two klystrons at the main LIGHT frequency of 2997.92 MHz. Passing through the four SCDTL cavities, the beam will accelerate to 37.5 MeV. A SCDTL RF unit consists of a LLRF driving power to a Modulator Klystron System (MKS) that amplifies the 5 microseconds duration RF pulses to MW levels. From each SCDTL RF unit, the associated LLRF receives 4 probe signals (2 per cavity) and 3 pairs of directional couplers (forward and reflected) signals in the RF network. The power is split from main line and there is a coupler on each branch before the SCDTL cavities.
- Fifteen CCL (Coupled Cavity LINAC) structures powered by eleven klystrons at 2997.92 MHz, bringing the beam energy up to 230 MeV. Four CCL RF units split power between two CCL cavities and the other six are fed directly from the MKS. In the first case, the associated LLRF receives 4 probe signals (2 per cavity) and 3 pairs of directional couplers (forward and reflected) signals in the RF network (as in the SCDTL case); and in the latter case only 2 probe signals from the cavity and 2 pairs of directional couplers are received.

Figure 1 shows a schematic view of the LIGHT LINAC design. The three types of RF accelerating cavities are highlighted with the final energy at each section. The RF peak power required for the RFQ, SCDTL and CCL units is 400 kW, 8 MW and 45 MW respectively [2].

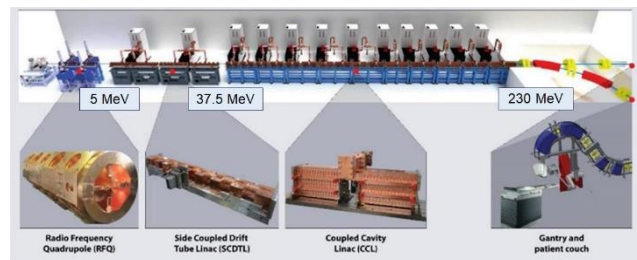

Figure 1: LIGHT system schematic with expected beam energy reached after each type of RF accelerating cavity section.

The LIGHT beam production system is currently being commissioned at AVO-ADAM Daresbury integration site (DIS) in UK.

## LLRF SYSTEM DESCRIPTION

Each high-power RF unit is fed by a LLRF device, which makes a total of 14 LLRF units for the whole LIGHT LINAC. The LLRF system has been built and delivered by Instrumentation Technologies [2] and on top of the LLRF device units, it's composed of a Reference Master

_________________________
† d.soriano@avo-adam.com

Oscillator (RMO) at 2997.92 MHz and its distribution box (to amplify signal or use a sub harmonic frequency), and an interlock unit box. The interlock unit box receives the accelerator control system interlock signals and sends them to each LLRF box.

A 3 GHz LLRF device has a front-end unit and a digital processor unit. The former is responsible of acquiring RF signals and passing them at the intermediate frequency (IF) to the digital processor, where they are checked and sent to the accelerator control system. The digital processor also shapes the pulse in phase and amplitude (as requested for each treatment energy) and sends this information back to the front-end where the RF output signal is delivered. More information can be found in [2].

## LLRF SYSTEM FEATURES

The function of the LLRF system in the LIGHT LINAC is two-fold: produce RF signal that will be amplified and transported to the RF cavities and acquire RF signals from the different probes and directional couplers.

### RF stability

Small energy spread in the beam requires stringent pulse-to-pulse stability of the LLRF RF output in amplitude and phase. In Figure 2 and 3 an example of a 500-pulse run is shown for both amplitude and phase sampled at the pulse flattop. The RMS deviation in amplitude is 0.013% and in phase 0.011 deg.

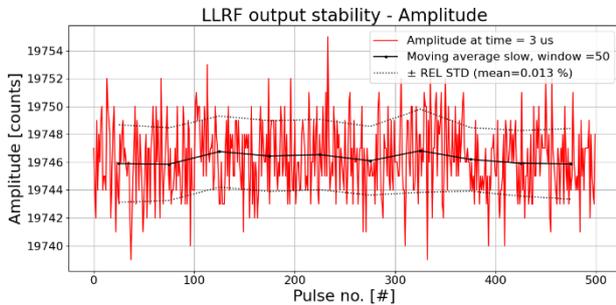

Figure 2. LLRF output amplitude (red) measured for 500 pulses with a moving average (solid black) and standard deviation (dotted).

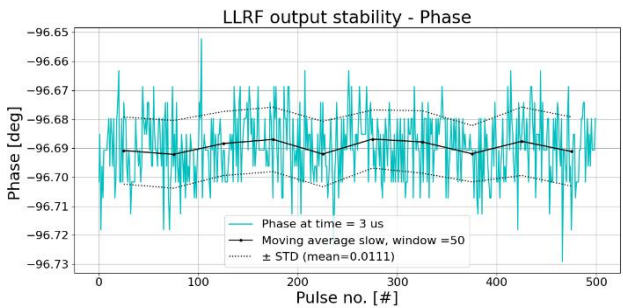

Figure 3. LLRF output phase (cyan) measured for 500 pulses with a moving average (solid black) and standard deviation (dotted).

### Acquisition and feedback loop

An RF probe signal is used in a feedback loop to maintain amplitude and phase at values set by operators. Amplitude and phase feedback loops have independent PI controllers. An example of the feedback loop behaviour from the RFQ commissioning tests is shown in Figure 4. Before the loop was enabled (blue), the water temperature (green) was oscillating and so was the RFQ amplitude (red). Once enabled, around 18:05, the RFQ amplitude became much more stable for each setpoint, not affected by water temperature oscillations.

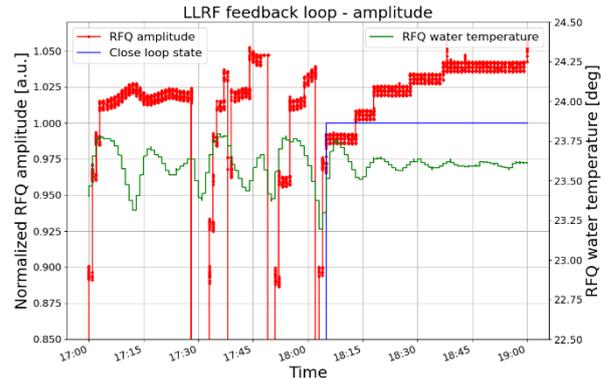

Figure 4. LLRF amplitude feedback loop, set on at 18:05 (blue), and its effect on RFQ amplitude (red). Also shown RFQ water temperature (green) oscillations causing drifts in amplitude.

### Pulse shaping

The LLRF can shape the output pulse in amplitude and phase defining the I and Q components through a series of second order splines. This allows having a LLRF output pulse that compensates for any flatness deviation coming from the high voltage pulse of the high-power amplification stations, for example. An RF output pulse with parabolic sections is shown in Figure 5. Up to 25 sections can be defined, providing great flexibility to compensate for any non-flatness.

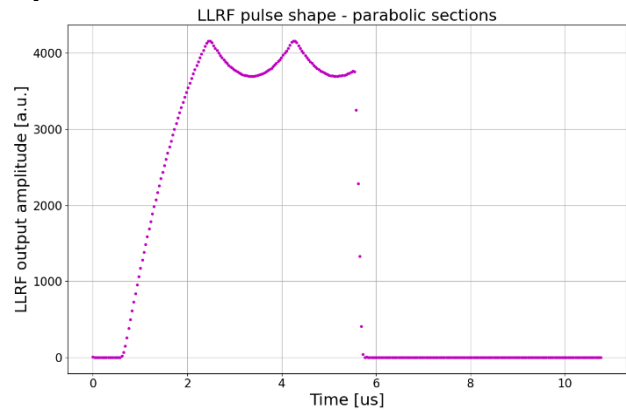

Figure 5. LLRF output pulse shape with 3 parabolic sections defined.

### Breakdown detection and resonance frequency feedback

During first stages of RF conditioning and while ramping up RF power, breakdown inside the RF cavities can develop and cause damage to the inner surfaces. Therefore, the LIGHT LLRF system was equipped with a breakdown detection functionality capable of performing a check on every pulse to ensure the ratio between forward and reflected signals (from directional couplers) is below an expert defined ratio. If the ratio exceeds this threshold, an event is counted and after a user-defined number of events per unit time, the LLRF will stop sending RF and notify the control system of its failure state.

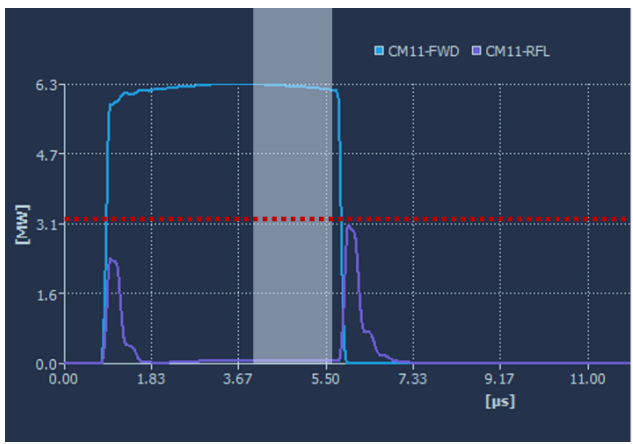

Figure 6. Forward and reflected RF signals from CCL module 11. A breakdown is counted if the reflected (purple) goes above a certain ratio (red dotted line at 50%, for example) of the forward (blue) within a user defined time gate.

Breakdowns occurrence is high during RF conditioning and automation of the process has been proven to be effective in preventing the cavities from experiencing uncontrolled level of breakdowns or breakdown clusters [3].

In addition to this feature, LLRF was required to provide a frequency error based on the analysis of the probe signal. The frequency error is sent to cooling control system to adjust the temperature of cooling water for that cavity to minimize the error.

### Front-end controller

A front-end controller (FEC) was put in place by AVO-ADAM to control and interact with the LLRF devices. The FEC hosts several interfaces:

- Data streaming interface: a unidirectional channel to receive pulse-related measurements.
- Real-time interface: a bidirectional deterministic request-response interface to configure each pulse according to the required treatment energy and receive the measured signals before the next pulse (within 5 milliseconds).
- Slow control interface: a bidirectional channel to control, configure each LLRF device and collect acquisition data every 100 milliseconds.
- Trigger interface: a unidirectional channel to control the trigger signal delivery. Each LLRF device has two trigger inputs: one for treatment/beam operation and another for only RF operation, with independent amplitude, phase, and pulse shape configurations.

In addition, the FEC performs a pulse shape analysis on all digitized waveforms provided by the LLRF device and makes both the digitized and the calculated data available to the upper-tier control system for live monitoring and for further processing.

## INSTALLATION DESCRIPTION AND STATUS

All LLRF units are mounted and distributed in four 42U racks on the klystron gallery at DIS. Currently all the 14 LLRF units have been installed and have passed their Site Acceptance Test (SAT). The Instrumentation Technologies LLRF platform allows to automatize a large part of the SAT process to facilitate and speed up the work of the engineers onsite. After the SAT is completed the LLRF unit is used for the high-power stations' RF SAT (standalone test to check their RF requirements), and for the calibration/integration tests aiming to have it configured correctly for the next beam commissioning phase.

### LLRF calibration

To provide accurate readings of the RF power from the accelerating cavities and RF network directional couplers, the LLRF input signals must be calibrated. The calibration consists in finding a linear coefficient to convert the number of counts read by the LLRF Analog to Digital Converters (ADC) into a real RF voltage/power. This is accomplished by considering RF line (cables, adaptors, rack patch panels) losses from the probe, and the coupling of the latter (same for the directional couplers). Once known, a controlled and calibrated RF source is fed to each LLRF ADC port to measure the response in ADC counts for each power level within the dynamic range (up to 20 dBm). An automated procedure collects the data and calculates the calibration coefficients for each port with expected errors from the instrument's uncertainty.

From our experience, the most difficult aspect of the calibration is the reliability of the measurements done in low power due to potential changes in the high-power part of the RF signal chain (thermal effects). However, for beam acceleration purposes it's not critical since the beam energy is the final observable.

## CONCLUSION

The LIGHT LLRF system was produced and delivered by Instrumentation Technologies to AVO-ADAM, and it has been fully SAT tested in Geneva and Daresbury for the LIGHT proton therapy LINAC commissioning. The performance was found within the specifications of the LIGHT requirements. The system was successfully deployed for high RF power station SATs, cavity conditioning, and 230 MeV beam commissioning [4].

## ACKNOWLEDGEMENTS

The authors would like to acknowledge the tempestive and professional support of the Instrumentation Technology team during the tests which allowed us to perform the deployment and validation of the AVO-ADAM LLRF system in a timely manner.